\documentclass[aps,prb,twocolumn,superscriptaddress,citeautoscript]{revtex4-1}

\usepackage{amsmath,amssymb}
\usepackage{bm}
\usepackage{graphicx}
\usepackage{epstopdf}
\usepackage{latexsym}
\usepackage{subfigure}
\usepackage{color}
\usepackage{dsfont} 
\usepackage{wasysym} 

\newcommand{\ncfo}{{NaCaCo$_2$F$_7$}}

\begin{document}

\title{Static and dynamic $XY$-like short-range order in a frustrated magnet with exchange disorder}

\author{K.A. Ross}
\altaffiliation[Current Address: ]
    {Colorado State University, Fort Collins, Colorado, 80523}
\affiliation{Institute for Quantum Matter and Department of Physics and Astronomy, Johns Hopkins University, Baltimore, Maryland 21218, USA}
\affiliation{NIST Center for Neutron Research, National Institute of Standards and Technology, Gaithersburg, Maryland 20899, USA}

\author{J.W. Krizan} 
\affiliation{Institute for Quantum Matter, Princeton University, Princeton, New Jersey 08544, USA}

\author{J.A. Rodriguez-Rivera} 
\affiliation{NIST Center for Neutron Research, National Institute of Standards and Technology, Gaithersburg, Maryland 20899, USA}
\affiliation{Materials Science and Engineering, University of Maryland, College Park, MD 20742, USA}

\author{R.J. Cava} 
\affiliation{Institute for Quantum Matter, Princeton University, Princeton, New Jersey 08544, USA}

\author{C.L. Broholm} 
\affiliation{Institute for Quantum Matter and Department of Physics and Astronomy, Johns Hopkins University, Baltimore, Maryland 21218, USA}
\affiliation{NIST Center for Neutron Research, National Institute of Standards and Technology, Gaithersburg, Maryland 20899, USA}
\affiliation{Department of Materials Science and Engineering, The Johns Hopkins University, Baltimore, Maryland 21218, USA}

\date{\today}
\begin{abstract}
 A single crystal of the Co$^{2+}-$based pyrochlore \ncfo \ was studied by inelastic neutron scattering.  This frustrated magnet with quenched exchange disorder remains in a strongly correlated paramagnetic state down to one 60th of the Curie-Weiss temperature.  Below $T_f = 2.4$ K, diffuse elastic scattering develops and comprises $30 \pm 10\%$ of the total magnetic scattering, as expected for $J_{\text{eff}} = 1/2$ moments frozen on a time scale that exceeds $\hbar/\delta E$=3.8 ps.   The diffuse scattering is consistent with short range $XY$ antiferromagnetism with a correlation length of 16 \AA.  The momentum ($\boldsymbol{Q}$) dependence of the inelastic intensity indicates relaxing $XY$-like antiferromagnetic clusters at energies below $\sim$ 5.5 meV, and collinear antiferromagnetic fluctuations above this energy.    The relevant $XY$ configurations form a continuous manifold of symmetry-related states.  Contrary to well-known models that produce this continuous manifold, order-by-disorder does not select an ordered state in \ncfo \ despite evidence for weak ($\sim 12 $\%) exchange disorder.  Instead, \ncfo \ freezes into short range ordered clusters that span this manifold. 
\end{abstract}

\maketitle

\section{Introduction}
The spin liquid state of the Heisenberg antiferromagnet (HAFM) on the pyrochlore lattice supports fluctuations within an extensively degenerate ground state manifold consisting of correlated, yet disordered, spin configurations \cite{moessner1998low,moessner1998properties,lacroix2011introduction}.   This beautiful state of matter arises from a perfect frustration of antiferromagnetic (AFM) interactions on the corner sharing tetrahedra that comprise the pyrochlore lattice.  However, the spin liquid is extremely susceptible to small perturbations that can reduce the ground state degeneracy and lower the free energy.   The manner in which the spin liquid is modified in real materials with deviations from ideal Heisenberg exchange is thus a rich field of study, with many possible outcomes depending on the relevant perturbations \cite{gardner2010magnetic}.  In particular, the role of fluctuations in selecting subsets of the ground state manifold must often be considered.  Thermal and quantum fluctuations that are softer for certain spin configurations can, in some cases, select long range ordered (LRO) states in a mechanism called order-by-disorder \cite{champion2003er, zhitomirsky2012quantum, savary2012order, maryasin2014order, wong2013ground, mcclarty2014order}.

Quenched disorder, in the form of vacancies or bond disorder (i.e., local variations in the strength of the spin-spin interactions), also produces order-by-disorder, as described in the pioneering work by Villain \cite{villain1980order} and later studied in detail by others \cite{henley1987ordering, henley1989ordering,maryasin2014order,mcclarty2014order}.  Quenched disorder can compete with thermal fluctuations to determine the ordered state.  An important recent example is the $XY$ antiferromagnetic pyrochlore material Er$_2$Ti$_2$O$_7$.   For the pseudospin $\frac{1}{2}$ model believed to be appropriate for this material, thermal and quantum order-by-disorder have been shown to select a non-coplanar LRO state \cite{zhitomirsky2012quantum,savary2012order, oitmaa2013phase}, while quenched disorder is predicted to favor a coplanar LRO state in the same model \cite{maryasin2014order, andreanov2015order}.  The role of quenched disorder for the HAFM pyrochlore model has been studied in the past by including a distribution of exchange interactions spanning $\bar{J} \pm \Delta$ in the HAFM Hamiltonian, $ H = \sum_{ij} J_{ij} 
\mathbf{S}_i\cdot \mathbf{S}_j$.  In the limit of weak disorder, $\Delta << \bar{J}$, where $\bar{J}$ is the mean exchange interaction, the spins are expected to form locally collinear antiferromagnetic correlations and the system eventually freezes at a temperature $T_f \approx \Delta$ \cite{saunders2007spin,andreanov2010spin,bellier2001frustrated}.

\begin{figure}[!tb]  
\centering
\includegraphics[ width=0.9\columnwidth]{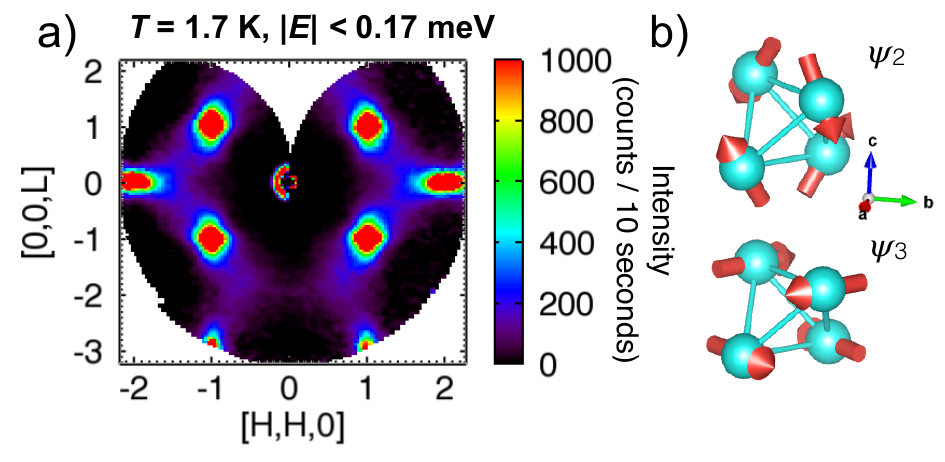}
\caption{ a) Elastic magnetic scattering from \ncfo \ obtained by subtracting elastic scattering at $T$ =14 K from that at 1.7 K.   Two diffuse components are visible: broadened Bragg peaks occurring at nuclear-allowed positions, and an extended ``zig-zag'' component that is also apparent at finite energy transfers [Fig. \ref{fig:fig3} b)].  The data were acquired at fixed monitor counts; we report approximate counting time per setting (10 seconds).  The total time required to collect the data shown here was 32 hours, with 16 hours per temperature.  b) The two basis states of the $\Gamma_5$ IR ($XY$ states), combinations of which form the short range order in \ncfo.}
\label{fig:fig1}
\end{figure}

Here we report on the nature of static and dynamic spin correlations in the recently discovered pyrochlore material, \ncfo, which has been synthesized in single crystal form via the optical floating zone method  \cite{krizan2014nacaco}.  The inherent local disorder arising from the mixed-charge $A$-site (Na$^{+}$/Ca$^{2+}$) is expected to yield exchange disorder.  The availability of large single crystals has allowed us to measure the full dynamic structure factor for \ncfo \ in the high symmetry [$HHL$] plane.  Our measurements indicate \ncfo \ adheres to the expectations for the HAFM with weak exchange disorder in the high energy limit ($E > 5.5$ meV), but at low energies it displays frozen short range correlations and relaxational dynamics associated with an easy plane ($XY$-like) manifold.  

In \ncfo \ (space group $Fd\bar{3}m$, $a$ =10.4056(2) \AA \ at $T$ = 295 K \cite{krizan2014nacaco}) the $A$-site of the pyrochlore lattice is occupied by Na$^{+}$ and Ca$^{2+}$ ions with equal concentration in a disordered configuration.  Thermodynamic magnetic properties of \ncfo \ evidence a spin-freezing transition at $T_f \approx 2.4$ K \cite{krizan2014nacaco}.   The low freezing temperature indicates \emph{weak} exchange disorder with $\Delta/\bar{J} \sim 0.12$ (assuming that $T_f \approx \Delta$, as in the models of Refs. \onlinecite{saunders2007spin, andreanov2010spin} and $\bar{J} \sim 20 $ K based on the Curie-Weiss temperature, $\theta_{CW}$ = -140 K, and assuming $S=3/2$).   The effective moment determined from Curie-Weiss analysis at high temperatures is 6.1 $\mu_B$; this large effective moment obtained at high temperatures indicates significant thermal population of a $J_{\text{eff}}$ = 3/2 quartet at room temperature.  Nonetheless, the change in entropy at low temperatures approaches $R\ln2$ \cite{krizan2014nacaco}, suggesting a spin-orbit coupled ground state Kramers doublet with $J_{\text{eff}} = 1/2$, as is often relevant to Co$^{2+}$ materials \cite{maartense1977field, regnault1977magnetic, zhou2012successive, kenzelmann2002order}.   The degree of anisotropy of the effective moments in NaCaCo$_2$F$_7$ is not yet determined, but magnetization at $T=2$ K and 40 K is linear and isotropic up to $\mu_0 H =  9$ T (Appendix \ref{sec:magnetization}) \cite{krizan2014nacaco}.

 \section{Experimental Method}
We studied a 0.87 g single crystal of \ncfo \ using the MACS spectrometer at the NIST Center for Neutron Research \cite{rodriguez2008macs}.  The dynamic structure factor, $S(\mathbf{Q},E)$, was measured in the [$HHL$] reciprocal lattice plane.  Two configurations were used for spectroscopic measurements; for low energy transfer scans, neutrons with a final energy $E_f = 3.7$ meV were selected, and post-sample BeO filters were used to remove higher harmonic contamination and reject neutrons for which $E_f>3.7$ meV.  For incident energies below (above) $E_i$ = 5.2 meV, a Be filter (open channel) preceded the sample.  For energy transfers above $E=8.2$ meV, fixed  $E_f = 5.0$ meV was used with post-sample Be filters and an open channel pre-sample.  In all figures, measurements made in the various configurations are normalized to the same intensity units (counts per monitor units) using overlapping energy scans.   The energy resolution was $\delta E$ = 0.17 meV at the elastic line for $E_f = 3.7$ meV and $\delta E$ = 0.34 meV for $E_f = 5.0$  meV.

\section{Results}
Below $T_f$, the elastic magnetic scattering indicates static short range AFM spin correlations.  The subtraction of 14 K from 1.7 K data reveals strong magnetic diffuse scattering (Fig. \ref{fig:fig1} a).  There are two components to the this diffuse scattering; diffuse Bragg spots (not resolution limited in $\boldsymbol{Q}$, as will be discussed below), in addition to diffuse scattering taking the shape of a ``zig-zag'' pattern underlying the peaks, i.e. the extended diffuse intensity is strongest along the lines connecting certain zone centers.  This zig-zag diffuse pattern persists at finite energy transfers (Fig. \ref{fig:fig3}), as will be discussed below.

The diffuse Bragg spots arise at the (111), (220), and (113) zone centers, but importantly \emph{not} at  (002) or (222) (see also Fig. \ref{fig:rawdata} in Appendix \ref{sec:moreneut}).  These absences strongly constrain the frozen spin configuration.   Although a short range structure based on collinear antiferromagnetic moments might be expected based on the weak disorder HAFM model, the absence of the (002) diffuse Bragg spot rules out this scenario.   The observed magnetic peaks are instead consistent with $XY$ spin configurations, specifically those transforming as the $\Gamma_5$ irreducible representation (IR) of the tetrahedral point group $T_d$.  The $\Gamma_5$ IR admits a continuous manifold of states parameterized by a single angular parameter, $\alpha$, which rotates the spin on each sublattice around its local $<111>$ axis (Appendix \ref{sec:gamma5})\cite{champion2003er, zhitomirsky2014nature}.   

\begin{figure}[tb!]  
\centering
\includegraphics[ width=\columnwidth]{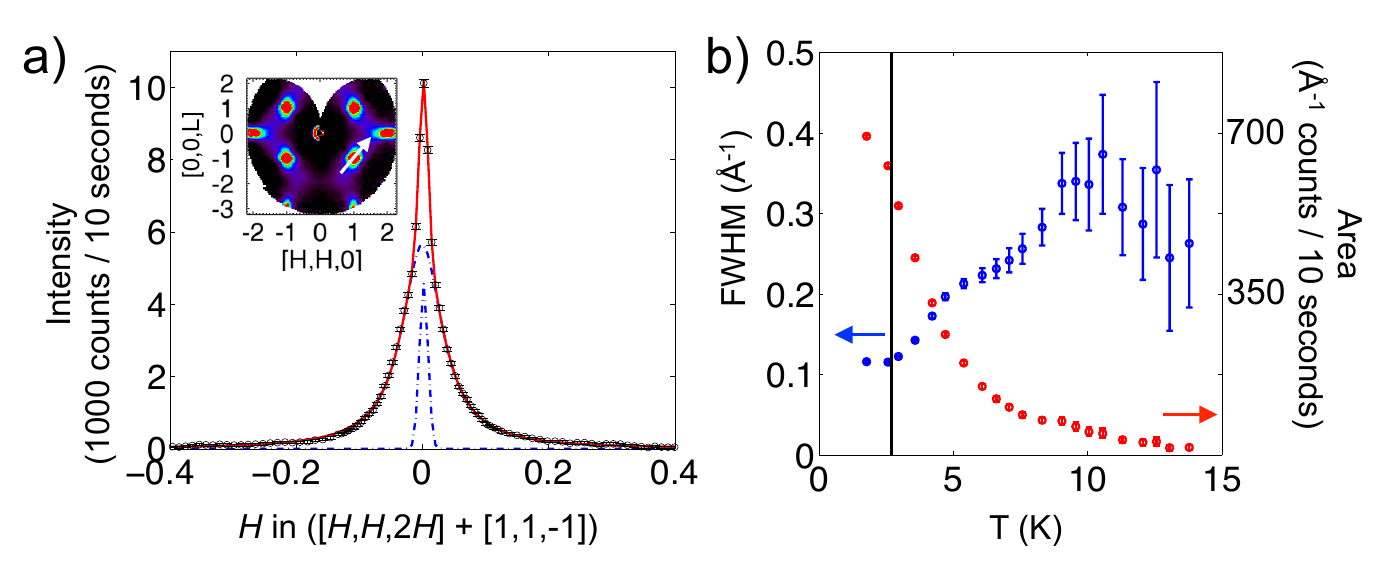}
\caption{a) Elastic scattering (1.7 K) transverse to the $(11\bar{1})$ position (scan direction shown in inset).  Solid line is a fit to a Gaussian (nuclear Bragg peak) plus Lorentzian (magnetic diffuse scattering).  The data were acquired at fixed monitor counts; we report approximate counting time per setting.  b) The temperature dependence of FWHM and area of the Lorentzian component of the elastic transverse scan.  The fitting parameters for the Gaussian were fixed to their values at 14 K.   Error bars represent one standard deviation.}
\label{fig:fig2}
\end{figure}

At 1.7K, the elastic peak near $(11\bar{1})$ (Fig. \ref{fig:fig2} a)) can be described by the sum of a sharp Gaussian and a broad Lorentzian component.  The Gaussian persists at all measured temperatures; this nuclear Bragg peak is a measure of the instrumental $\boldsymbol{Q}$-resolution.  The Lorentzian magnetic component gradually develops upon cooling from 14 K (Fig. \ref{fig:fig2}b), while the Full Width at Half Maximum (FWHM) decreases.  This behavior is reminiscent of critical scattering preceding a transition to an ordered state.  From this perspective, the transition in \ncfo \  may be thought of as being preempted by freezing.  Below $T_f$, the FWHM of the Lorentzian saturates at 0.12(1) \AA$^{-1}$, implying a correlation length of 16.1(1) \AA \ for the short range magnetic order.     

\begin{figure}[!tb]  
\centering
\includegraphics[ width=0.85\columnwidth]{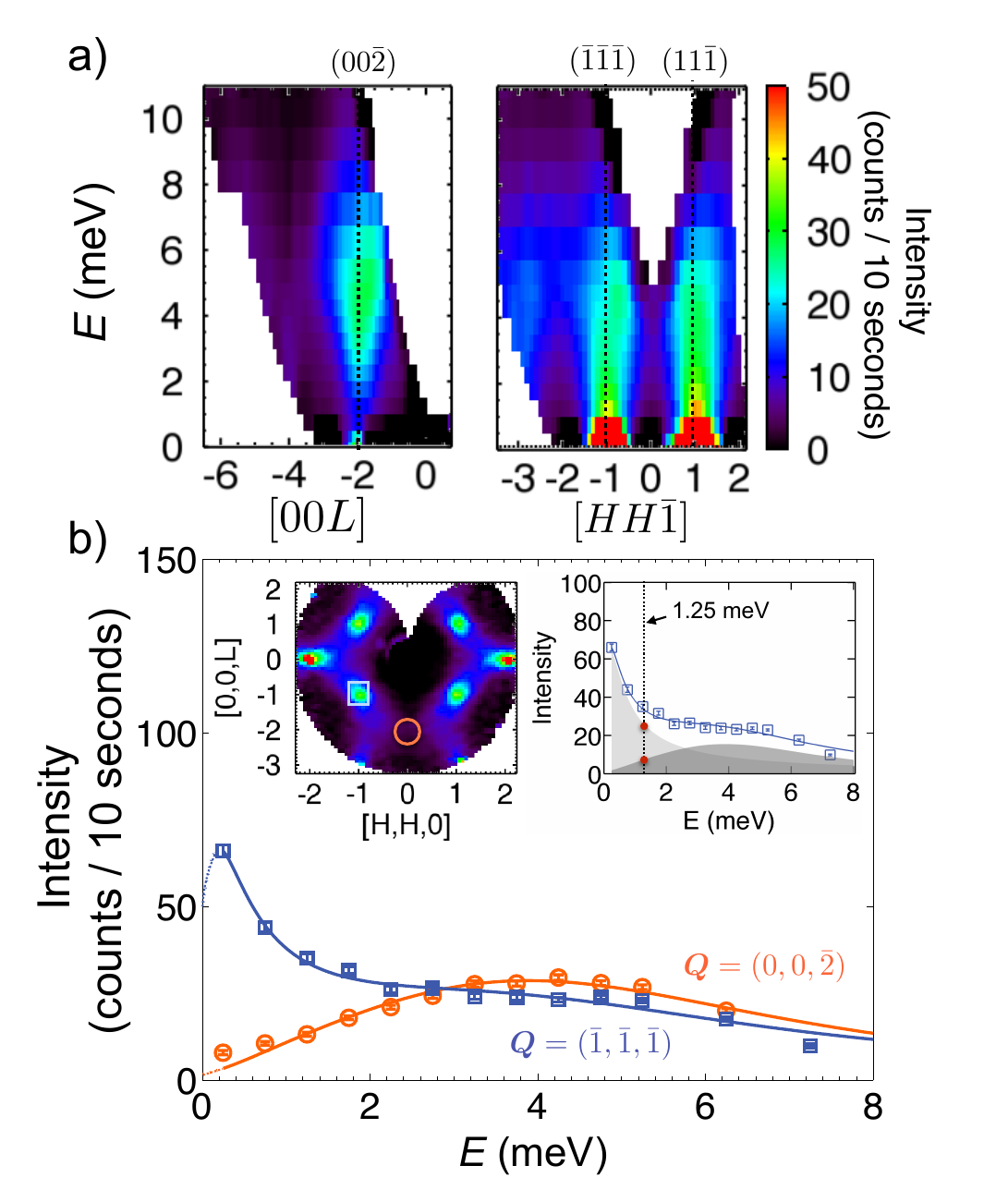}
\caption{ a) Magnetic scattering intensity versus energy and momentum transfer at $T$=1.7 K.  The inelastic intensity near $(00\bar{2})$ decreases as $E$ approaches zero (left panel), while it increases at the diffuse Bragg spots such as $(\bar{1}\bar{1}\bar{1})$ (right panel).  Averaging ranges: [$HH0$]:  $-0.1<H<0.1$ (left) and [$00L$]:  $-1.1<L<-0.9$ (right).   b) Cuts through $(\bar{1}\bar{1}\bar{1})$ and $(00\bar{2})$ (averaging $\pm$0.1 r.l.u. in $H$ and $L$).  The solid lines are fits, described in the main text. The left inset shows the location of the energy cuts on a constant $E=0.25$ meV map.  The right inset shows the decomposition of the $(\bar{1}\bar{1}\bar{1})$ fit into two components.  At 1.25 meV, the same energy as the slice shown in Fig. \ref{fig:fig3}b), both the DHO and Lorentzian components contribute substantially to the overall intensity (red circles). The data were acquired at fixed monitor counts; we report approximate counting time per setting. Errorbars represent one standard deviation.} 
\label{fig:fig4}
\end{figure}

\begin{figure}[!tbp]  
\centering
\includegraphics[ width=\columnwidth]{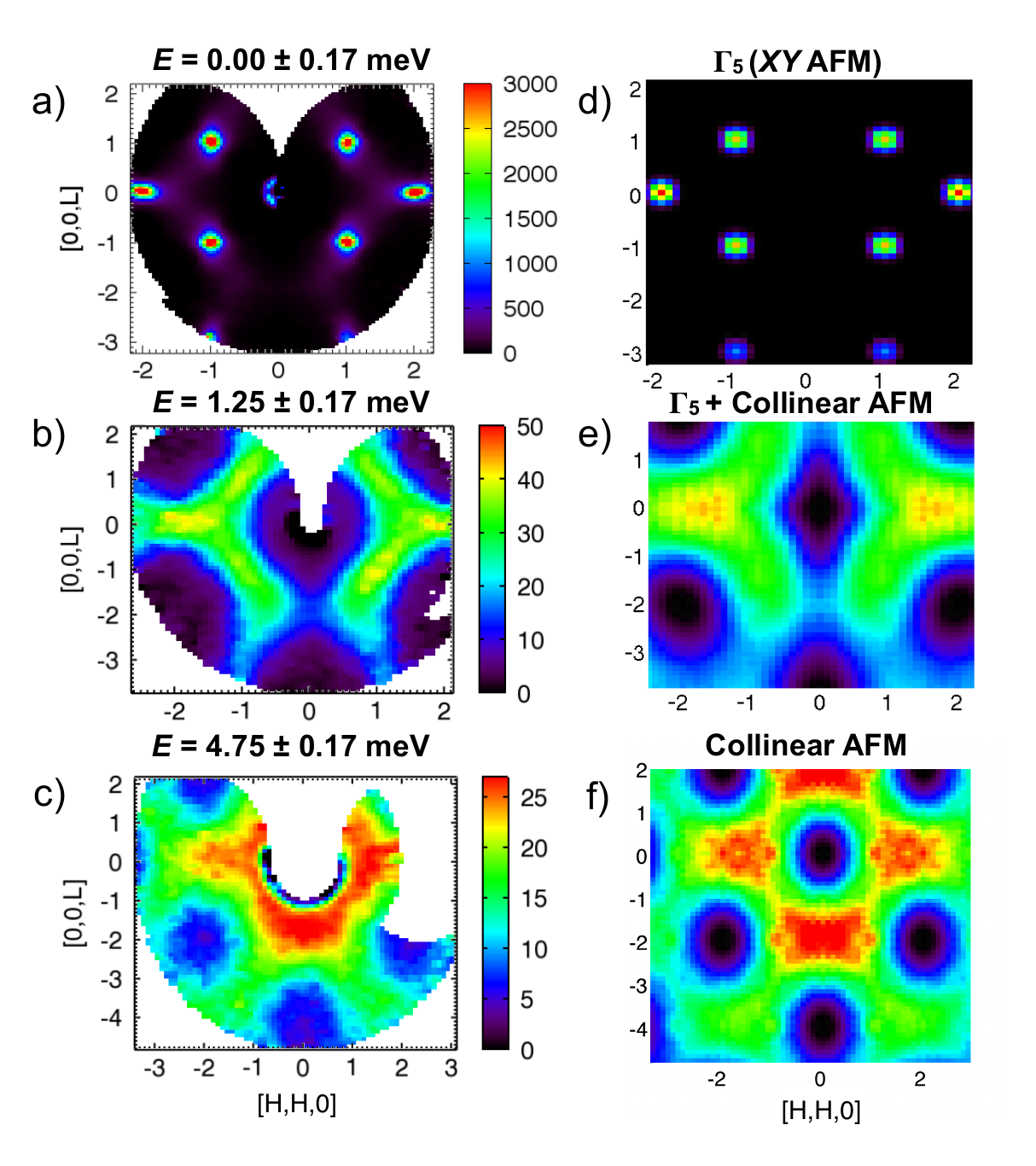}
\caption{Comparison of measured scattering, in counts per $\sim$10 seconds, at $T$=1.7 K (a,b,c) to simulated  patterns involving $XY$ and collinear AFM static spin configurations (d,e,f).  Measurements show diffuse scattering at a) $E = 0.00 \pm 0.17$ meV [14 K data subtracted, see also Fig. \ref{fig:fig1}], b) $E = 1.25 \pm 0.17$ meV, c) $E = 4.75 \pm 0.17$ meV.  Calculations of magnetic scattering from d) short range correlated cluster of $XY$ AFM order with symmetry $\Gamma_5$ and correlation length 16 \AA, e) independent tetrahedron model of 50\% $\Gamma_5$ states plus 50 \% collinear AFM spin configurations, and f) independent tetrahedron collinear AFM model.}
\label{fig:fig3}
\end{figure}

Inelastic scattering is also readily observed in \ncfo \ (Fig. \ref{fig:fig4}).   Using the total moment sum rule for magnetic neutron scattering, we find that the ratio of elastic magnetic to total magnetic scattering is $r = 0.3(1)$ in the measured region of the $[HHL]$ plane.  This is consistent with the ratio expected for a fully frozen (static) configuration arising from moments with $J_{\text{eff}}=\frac{1}{2}$ \big(for which $r = \frac{{J_{\text{eff}}}^2}{J_{\text{eff}}(J_{\text{eff}}+1)} = \frac{1}{3}$\big) but it is half of what is expected for bare $S=\frac{3}{2}$ ($r =$ 0.6).

To interpret the inelastic magnetic neutron scattering we examine $\boldsymbol{Q}-E$ slices (Fig. \ref{fig:fig4} a), the energy dependence of constant-$\boldsymbol{Q}$ cuts (Fig. \ref{fig:fig4} b), as well as the $\boldsymbol{Q}$-dependence of constant-$E$ slices (Fig. \ref{fig:fig3}).  The latter reflect the spatial Fourier transform of spin correlations with a characteristic fluctuation frequency of $\omega = E/\hbar$.  From the $\boldsymbol{Q}-E$ slices of Fig. \ref{fig:fig4} a), the spectrum of magnetic scattering at the diffuse Bragg spots is seen to be gapless, while the spectrum at systematically absent magnetic Bragg spots (e.g. $(00\bar{2})$) is gapped and strongly damped.   

The constant-$E$ slices reveal that the zig-zag structure of diffuse scattering, which is present in addition to the diffuse Bragg spots on the elastic line, persists to finite energy transfers (Fig. \ref{fig:fig3}b)).  For a more detailed analysis of these distinct spectra, Fig. \ref{fig:fig4} b) shows constant-$\boldsymbol{Q}$ cuts at $T$= 1.7 K at two locations in the $[HHL]$ plane.  For $\boldsymbol{Q}=(00\bar{2})$ where there is no elastic magnetic peak (orange symbols in Fig. \ref{fig:fig4} b), the data can be fit to a Damped Harmonic Oscillator (DHO) spectral function with an excitation energy of 5.5(1) meV and a damping coefficient 8.7(4) meV (i.e., an overdamped mode, see Appendix \ref{sec:fits}); we later identify this mode with collinear excitations out of the easy plane manifold.   In contrast, the inelastic spectra at the diffuse Bragg positions (blue symbols in Fig \ref{fig:fig4}b) can be fit to the sum of a Lorentzian relaxation function at $E=0$ meV with a HWHM of 0.33(1) meV and a DHO with central energy and damping fixed to the values extracted at $(00\bar{2})$ (Appendix \ref{sec:fits}).  This decomposition of the line shape of the spectrum at $(\bar{1}\bar{1}\bar{1})$ is detailed in the right inset of Fig. \ref{fig:fig4}b).    

The zig-zag pattern formed in the $[HHL]$ plane by the quasi-elastic scattering (Fig.\ref{fig:fig3} b)) can be associated with low energy states related by easy plane spin rotations that span the $\Gamma_5$ manifold.  To establish this, we first compared the constant energy slice at 1.25 meV to the calculated neutron scattering intensity from a spatial average of independent $XY$ tetrahedra (Appendix \ref{sec:moreneut}).  In this approximation, each tetrahedron supports one choice from the continuous $\Gamma_5$ manifold.  This independent $XY$ tetrahedra model captures the lack of intensity near the $(00\bar{2})$ position and the general zig-zag shape of the diffuse scattering.  A better agreement is obtained, however, when collinear AFM spin components are added to each $XY$ tetrahedron on the level of $\sim$ 50\% (Fig. \ref{fig:fig3} e)).  This indicates that while the frozen state is $XY$-like, excitations for energy transfers beyond the freezing temperature involve both in- and out-of plane spin components.  In particular, the DHO mode at 5.5 meV arises from out of plane excitations while the quasi-elastic component is associated with easy plane excitations.  The inset to Fig. \ref{fig:fig4} b) shows contributions from both types of excitations at 1.25 meV.   Beyond 4 meV, the constant energy slices (Fig. \ref{fig:fig3} c)) are reproduced by fully collinear AFM configurations on independent tetrahedra (Fig. \ref{fig:fig3} f)).   An important shortcoming of this simple model is the independent tetrahedra approximation. Specifically, the widths of the measured diffuse scattering for the slow spin fluctuations at 1.25 meV is significantly sharper in $Q$-space than predicted (Appendix \ref{sec:moreneut}), and this is then evidence for inter-tetrahedron correlations.  The size of the correlated region inferred from fitting raw data is 7.9 \AA, which may be compared to the 3.65 \AA \ side length of the tetrahedron.  Even at  $T=14$ K the inverse correlation length of the inelastic scattering is approximately half of that expected for independent tetrahedra.

\section{Discussion}

The above analysis of the inelastic spectrum at $T=$ 1.7 K suggests there are two types of dynamics in this short-range correlated system.  The first is a relaxing process that is qualitatively consistent with local rotations of $XY$ spin clusters through the continuous  $\Gamma_5$ manifold with a relaxation time of $\tau_{XY} = 2.02$ ps.   The second is a short-lived ($\tau_{H} = 0.15$ ps) inelastic mode at 5.5 meV, which is qualitatively consistent with collinear antiferromagnetic tetrahedral fluctuations, as might be expected from the weak disorder HAFM. 

The $XY$ spin configurations relevant for \ncfo \ at low energies are already well-studied in the pyrochlore literature. As shown in Appendix \ref{sec:gamma5}, the $\Gamma_5$ IR can be decomposed into two basis vectors which have commonly been called $\psi_2$ (non-coplanar) and $\psi_3$ (coplanar)  (Fig. \ref{fig:fig1} b).  LRO states based on the $\Gamma_5$ manifold are known to be selected in the HAFM model upon inclusion of ``indirect'' Dzyaloshinskii-Moriya (DM) interactions \cite{elhajal2005ordering}, despite an accidental ground state degeneracy admitting all values of $\alpha$ at the mean field level.    The same continuously degenerate manifold is also present at the mean field level for the $XY$ AFM pyrochlore model \cite{champion2004soft, mcclarty2014order}, and the $XY$-like anisotropic exchange model proposed for Er$_2$Ti$_2$O$_7$ \cite{zhitomirsky2012quantum,savary2012order,maryasin2014order,andreanov2015order}.  In all cases, the $\Gamma_5$ degeneracy is lifted by \emph{disorder}, and a LRO state is selected.  The ``disorder'' can arise from thermal or quantum fluctuations, or quenched exchange disorder.   However, in \ncfo, despite a clear mechanism for weak bond disorder, the $\Gamma_5$ degeneracy is retained and explored by the system on short length scales and long time scales.  In this case, exchange disorder does \emph{not} lead to spin order, but instead to a a frozen spin configuration that appears to span the continuous $\Gamma_5$ manifold.

The microscopic reason for the stabilization of $XY$ spin configurations in \ncfo \ is not yet certain.  However, the spin orbit coupled $J_{\text{eff}} = 1/2$ state expected for Co$^{2+}$ in a distorted octahedral environment could lead to $XY$ anisotropy, either in the $g$-tensor or the exchange interactions, or both.  The central energy of the damped inelastic mode (5.5 meV) may be a measure of the strength of the anisotropy.  This should be investigated in the future through measurements of single-ion energy levels of \ncfo.  Additional open questions, aside from quantifying the single ion anisotropy, include whether orbital and lattice degrees of freedom are relevant to \ncfo \ as in the related spinel compound GeCo$_2$O$_4$ \cite{tomiyasu2011molecular}.

\section{Conclusions}
In summary, \ncfo \ is the first example of a new class of pyrochlore single crystals based on a structurally ordered magnetic $3d$ transition metal site in a varying local environment created by a disordered non-magnetic site \cite{krizan2015single, krizan2015nacani2f7}.  The disordered environment leads to weak disorder in the strong AFM interactions in \ncfo \ ($\theta_{CW}$ = -140 K), and ultimately a low temperature freezing transition at $T_f$ = 2.4 K.  We have observed $XY$ spin configurations forming a short range ordered state below $T_f$ with a correlation length of $\xi = 16$ \AA.   The low energy fluctuations away from this frozen state are gapless to within the energy resolution of our measurement (0.17 meV) and take on a distinctive diffuse pattern that suggests relaxation through a continuous manifold of local $XY$ states.   At higher energies, a strongly damped mode at 5.5 meV dominates the spectrum.  The associated Q-dependence of the scattering intensity isconsistent with collinear antiferromagnetic tetrahedral fluctuations, indicating an $XY$ anisotropy barrier of $\sim$ 64 K.   

The continuous manifold of $XY$ spin configurations present in \ncfo \ is known to collapse to an ordered state via order-by-disorder in models relevant to Er$_2$Ti$_2$O$_7$ as well as by DM interactions in the pyrochlore HAFM.  However, unlike the aforementioned theoretical predictions, quenched exchange disorder in \ncfo \ does \emph{not} lead to the selection of an ordered state, but instead a quasi-static disordered state.  Apart from the low energy fluctuations that appear to span the $\Gamma_5$ manifold, a prominent out of plane damped mode is observed with the same local structure as predicted for the Heisenberg model with weak exchange disorder.  An intriguing aspect of \ncfo \ is the potential for ice-like correlations on the Na$^{+}$, Ca$^{2+}$ disordered sublattice. Such correlated disorder might be necessary to explain why \ncfo \ fails to develop long range order. 

\begin{acknowledgments}
The authors gratefully acknowledge enlightening discussions with O. Tchernyshyov, J.T. Chalker, and J.W. Lynn.  KAR acknowledges the hospitality of Colorado State University during the writing of this manuscript, and the use of the SPINDIFF software package \cite{paddison2013spinvert}.  The bulk of the work was supported by the US Department of Energy, office of Basic Energy Sciences, Division of Material Sciences and Engineering under grant DE-FG02-08ER46544.   In particular this included the crystal growth activities and neutron scattering experiments. This work utilized facilities supported in part by the National Science Foundation under Agreement No. DMR-0944772.  KAR was partially supported by NSERC of Canada.
\end{acknowledgments}

\appendix

\section{Definition of states in the $\Gamma_5$ manifold}
\label{sec:gamma5}
The sublattices of the pyrochlore lattice are described by the following fractional coordinates:

\begin{eqnarray}
&&\mathbf{d}_0=\left(\frac{3}{8}, \frac{3}{8},  \frac{3}{8}\right),\quad \mathbf{d}_1=\left(\frac{3}{8},  \frac{1}{8},  \frac{1}{8}\right),\\
&&\mathbf{d}_2=\left(\frac{1}{8},\frac{3}{8},\frac{1}{8}\right),\quad\mathbf{d}_3=\left(\frac{1}{8},  \frac{1}{8},  \frac{3}{8}\right).
\end{eqnarray}

The moments (pseudovectors) forming the $\psi_2$ and $\psi_3$ bases of the $\Gamma_5$ representation are assigned to these sublattices as:

\begin{equation}
\vec{\psi_2}
\left\{\begin{array}{l}
\mathbf{\hat{s}}_0=(1,1,\bar{2})/\sqrt{6}\\
\mathbf{\hat{s}}_1=(1,\bar{1}, 2)/\sqrt{6}\\
\mathbf{\hat{s}}_2=(\bar{1},1, 2)/\sqrt{6}\\
\mathbf{\hat{s}}_3=(\bar{1},\bar{1},\bar{2})/\sqrt{6},
\end{array}\right.,
\quad
\vec{\psi_3}
\left\{\begin{array}{l}
\mathbf{\hat{s}}_0=(1,\bar{1}, 0)/\sqrt{2}\\
\mathbf{\hat{s}}_1=(1, 1, 0)/\sqrt{2}\\
\mathbf{\hat{s}}_2=(\bar{1},\bar{1},0)/\sqrt{2}\\
\mathbf{\hat{s}}_3=(\bar{1},1,0)/\sqrt{2}
\end{array}\right.,
\label{eqn:spins}
\end{equation}

\begin{figure}[!h]  
\centering
\includegraphics[ width=\columnwidth]{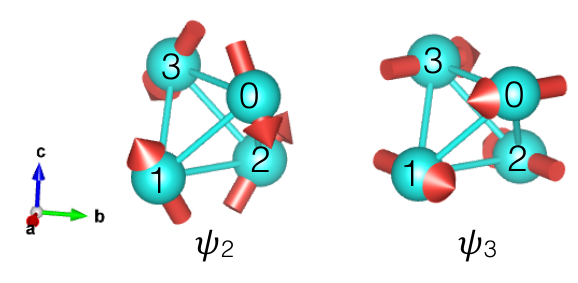}
\caption{Basis states of the $\Gamma_5$ IR (Eqn. \ref{eqn:spins})}
\label{fig:basisvectors}
\end{figure}

A general tetrahedral state with the symmetry of $\Gamma_5$ can be written as a linear combination of these sets,
\begin{equation}
\vec{\chi}(\alpha) = \cos{\alpha} \cdot \vec{\psi_2}  + \sin{\alpha} \cdot \vec{\psi_3}
\label{eqn:alpha}
\end{equation} 

\begin{figure*}[!htbp]  
\centering
\includegraphics[ width=2\columnwidth]{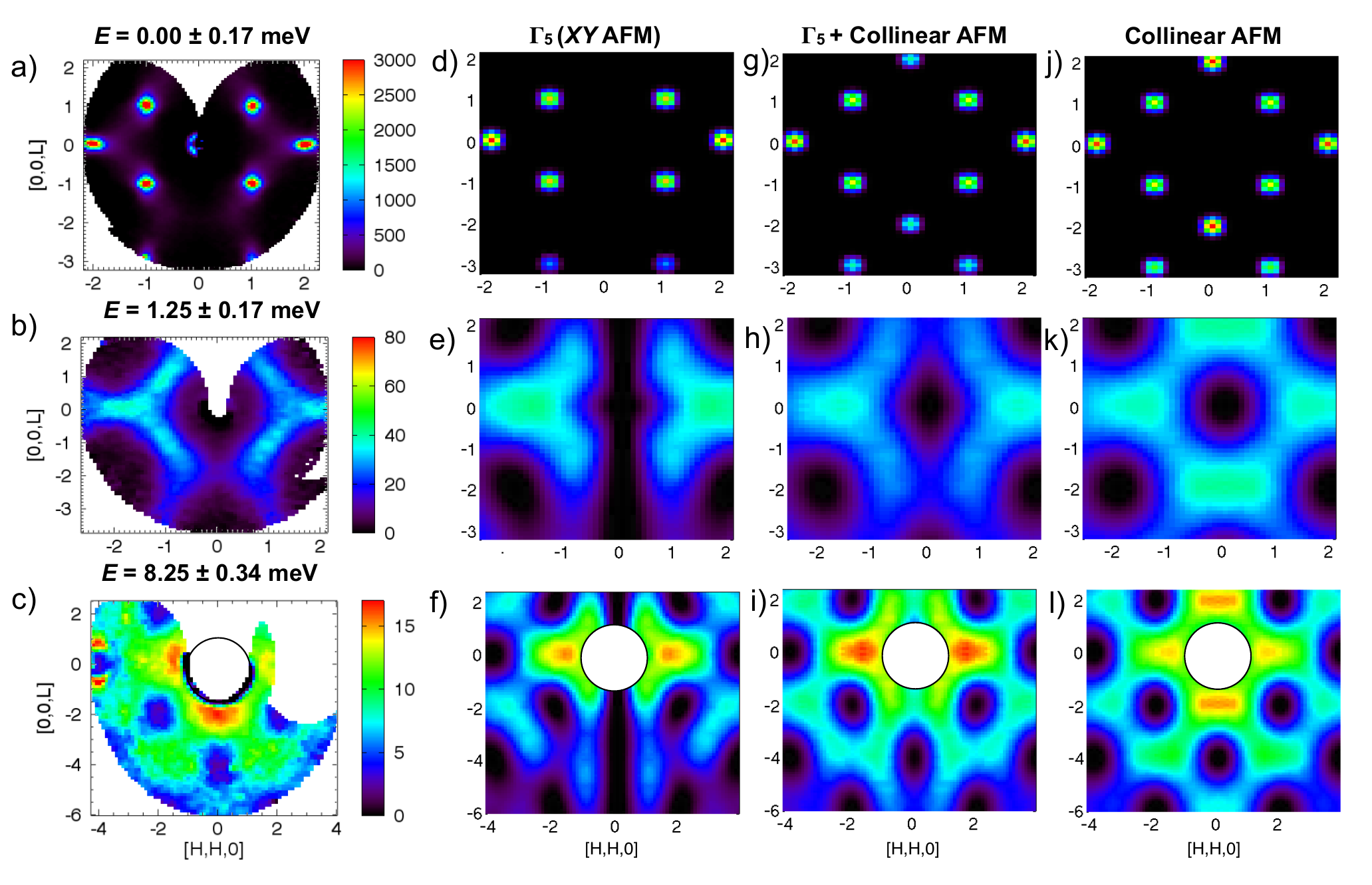}
\caption{Comparison of measured diffuse scattering (a,b,c) to calculations involving $XY$ (d,e,f), collinear AFM (j,k,l), and a mixture of the two types (g,h,i) of spin configurations.  Diffuse scattering at a) $E = 0.00 \pm 0.17$ meV, b) $E = 1.25 \pm 0.17$ meV, c) $E = 8.25 \pm 0.34$ meV.  d),e),f) calculations of magnetic scattering for short range correlated cluster of $\Gamma_5$ $XY$ AFM with correlation length 16 \AA \ (d) and independent $XY$ tetrahedra (e,f).  g),h),i) calculations involving tetrahedra with 50\% $XY$ and 50\% collinear AFM components [g) SRO with 16\AA, h)i) independent tetrahedra].  j),k),l) calculations involving collinear AFM configurations. Intensities and $Q$-ranges in f), i) and l) are rescaled from e), h), and k) to compare to panel c). For the independent tetrahedra models, each diffuse scattering pattern is computed using 50 instances of a 7$\times$7$\times$7 unit cell lattice (68600 tetrahedra in total).}
\label{fig:fullcomparison}
\end{figure*}

Assigning each ``up'' tetrahedron in the pyrochlore lattice a state $\vec{\chi}$ with a random value of $\alpha$ constitutes the independent tetrahedron $XY$ AFM state that is modeled in Figure \ref{fig:fullcomparison} e).  $\vec{\chi}(\alpha)$ spans a continuously deformable manifold of $XY$ states.  These are the relevant ground states at the mean field level for the case of Er$_2$Ti$_2$O$_7$ \cite{champion2003er, zhitomirsky2012quantum, savary2012order, maryasin2014order, wong2013ground, mcclarty2014order} or the HAFM model with ``indirect'' DM interactions \cite{elhajal2005ordering}.

\begin{figure}[!htbp]  
\centering
\includegraphics[ width=\columnwidth]{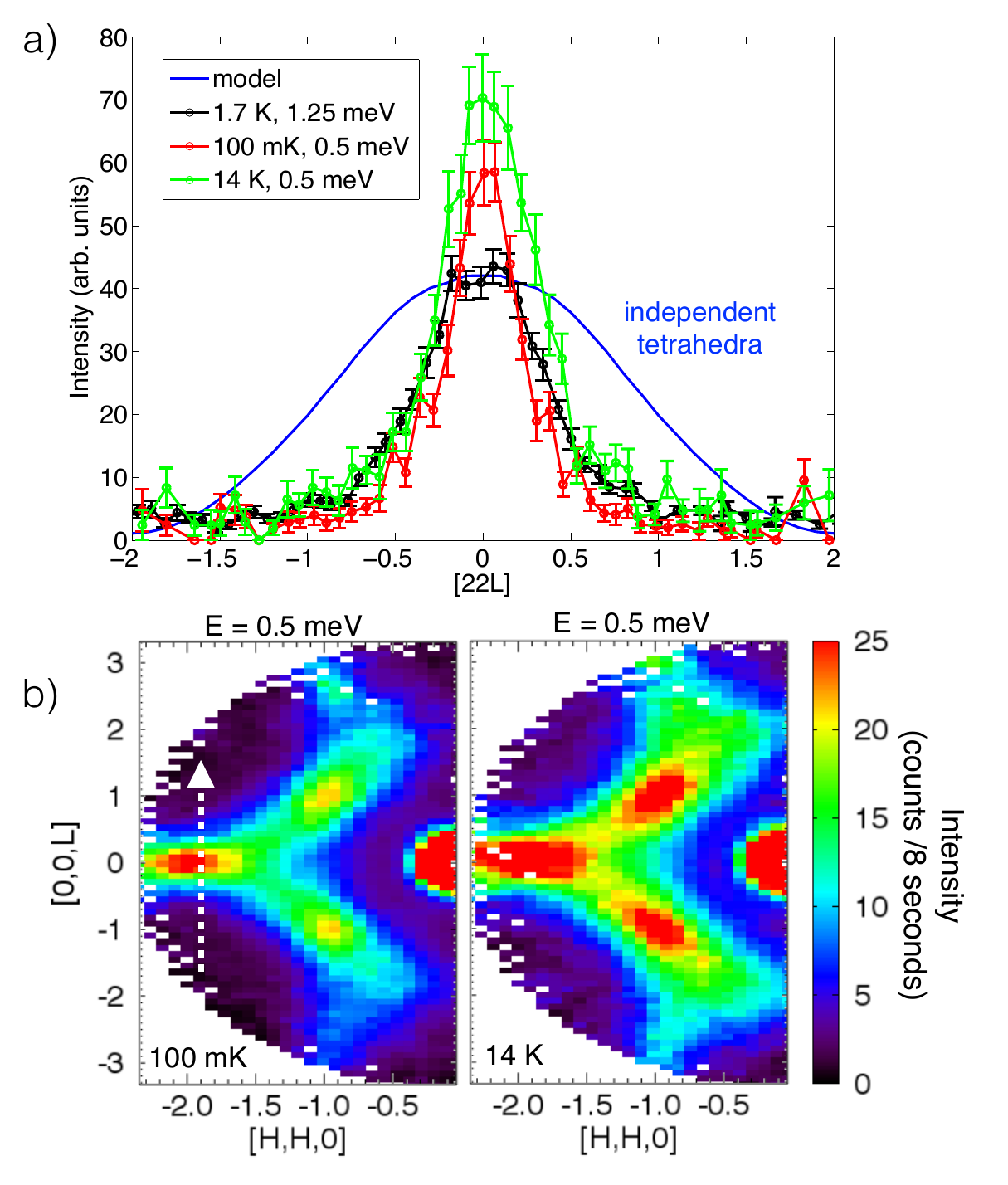}
\caption{ a) comparison of widths of $Q$ cuts near (220) along the [22L] direction, averaging over 0.28 r.l.u. in the perpendicular direction.  The black points are taken from the dataset presented in Fig. 4 b) of the main text.  The solid blue line is the width of the independent tetrahedron model represented in Fig. 4 e) of the main text.  Green and red points are cuts through the 0.5 meV data shown in panel b) of this figure.    b) inelastic intensity at $E=0.5$ meV at 100 mK (left) and 14K (right).  The zig-zag pattern in the 14 K (above $T_f$) data is broader and more intense than at 100 mK (below $T_f$).  White arrow shows location and direction of cuts in a).}  
\label{fig:widths}
\end{figure}

\begin{figure*}  
\centering
\includegraphics[ width=2\columnwidth]{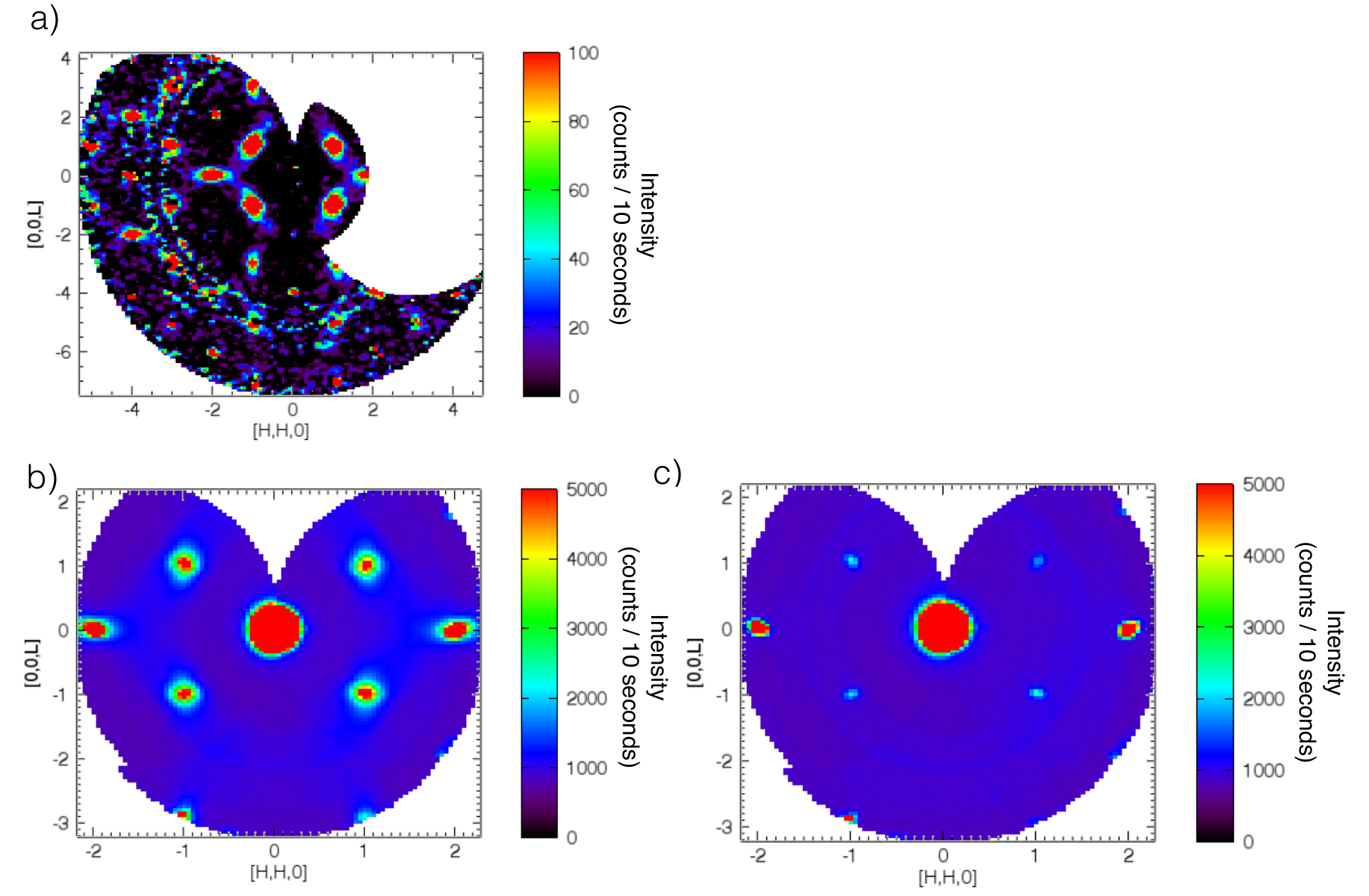}
\caption{ a) Elastic scattering at 1.7 K (14 K subtracted) taken with $E_i = E_f$ = 13.5 meV.  Pyrolytic graphite filters were used before and after the sample.  Rings are due to an imperfect subtraction of the aluminum powder lines arising from the sample mount.   b) Raw elastic data at 1.7 K (no subtraction) using $E_f = 3.7$ meV c) raw elastic data at 14K using $E_f = 3.7$ meV. }  
\label{fig:rawdata}
\end{figure*}

\begin{figure*}[!t]
\centering
\includegraphics[ width=2\columnwidth]{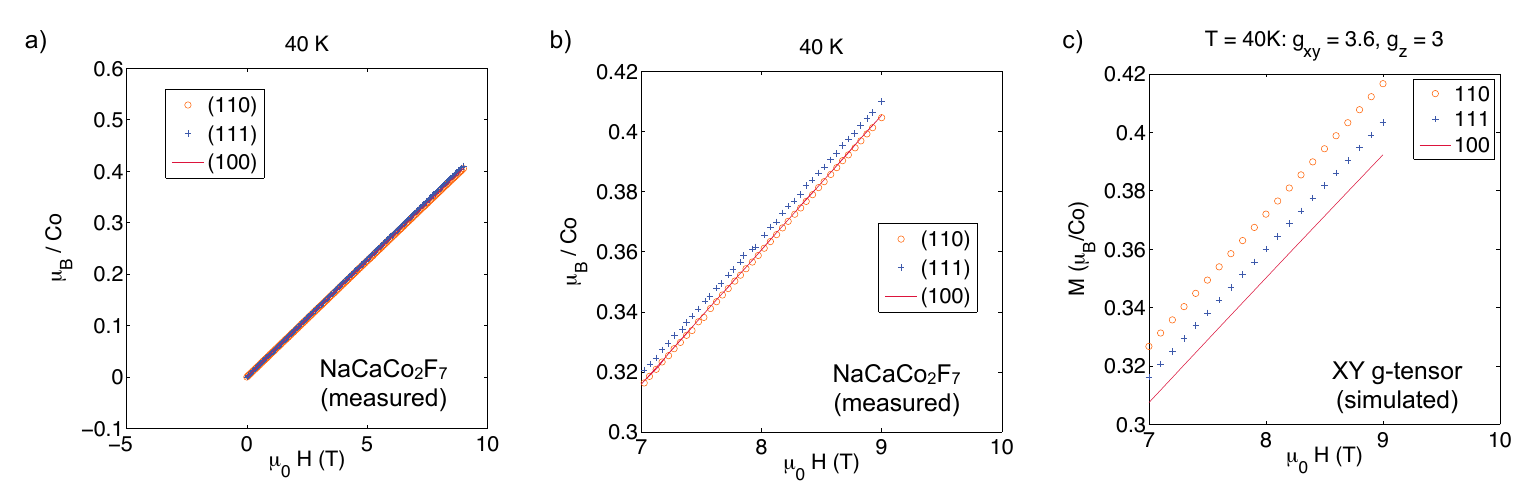}
\caption{Magnetization vs. magnetic field applied along three crystallographic directions. a), b) Magnetization data for a single crystal of \ncfo \ at $T=40$ K.  b) Slight differences in magnetization for the three directions can be seen a high fields, but the hierarchy of magnetization does not correspond to the expectations for $XY$-like or Ising-like g-tensors.  c) Calculation for an ideal pyrochlore paramagnet (Eqn. \ref{eqn:brillouin}) with $g_{xy} = 3.6$, $g_z = 3.0$, $J = 1/2$, and $T = 40$ K.}  
\label{fig:mag}
\end{figure*}

\section{Details of constant-$\boldsymbol{Q}$ lineshapes}
\label{sec:fits}

Figure 3 in the main text presents fits to constant-$\boldsymbol{Q}$ cuts, i.e. $S(E)$.  The fits to $S(E)$ include a relaxing diffusive component, $S_{XY}(E)$, and a damped harmonic oscillator, $S_{H}(E)$.  These have the well-known forms \cite{lovesey1984theorych5,lovesey1984theoryB},
\begin{equation}
S_{XY}(E) = \frac{A_{XY}E(1+n(E))}{\pi}\frac{\Gamma_{XY}}{E^2 + \Gamma_{XY}^2}, 
\label{eqn:Sxy}
\end{equation}
and,
\begin{eqnarray} 
 S_{H}(E) = A_H(1+n(E)) \times \frac{2\Gamma_H E}{(E^2-E_c^2)^2 + (2\Gamma_H E)^2}
\label{eqn:Sh}
\end{eqnarray}

where $\Gamma_{XY}$ is the HWHM of the diffusing component, and 2$\Gamma_{H}$ is the damping parameter of the DHO component.  The relaxation rates are then given by $\tau_{XY}$ = 1/$\Gamma_{XY}$ and $\tau_{H}$ = 1/$\Gamma_{H}$ (with $\Gamma$'s expressed in units of frequency).  $A_{XY}$ and $A_{H}$ are scale factors in arbitrary units.  $n(E)$ is the Bose-Einstein population factor, $n(E) = (\exp({E/k_B T}) - 1)^{-1}$.  $E_c/\hbar$ is the frequency of the DHO mode.  

\begin{table}[h]
\begin{tabular}{cccccc}
                         & $A_{XY}$ & $\Gamma_{XY}$ (meV) & $A_H$ & $\Gamma_H$ (meV) & $E_c$ (meV) \\  \cline{2-6} 
\multicolumn{1}{l|}{$(00\bar{2})$} & --    &      --   & 1157(52)   &     4.3(2)   &  5.5(1)   \\
\multicolumn{1}{l|}{$(\bar{1}\bar{1}\bar{1})$} &    109(2) &    0.33(1)     &   625(17)  &     4.3   &   5.5
\end{tabular}
\caption{Parameters for fits of constant energy scans at $\boldsymbol{Q}$ = $(00\bar{2})$ and $\boldsymbol{Q}$ = $(\bar{1}\bar{1}\bar{1})$ to Eqns. \ref{eqn:Sxy} and \ref{eqn:Sh}.}
\end{table}


\section{Supporting Neutron Scattering Data}
\label{sec:moreneut}
Here we present additional information supporting the conclusions from the main text.  Figure \ref{fig:fullcomparison} shows a more detailed comparison to three choices of models; short range ordered states with 16 \AA \ correlation lengths, and single tetrahedron states with $XY$ or locally collinear character, or a mixture of these.  Fig. \ref{fig:widths} a) shows the widths of diffuse features as compared to the independent tetrahedra model, as well as both lower temperature (100 mK) and higher temperature (14 K) inelastic scans.   The low energy diffuse inelastic scattering corresponds to a correlated region ($\sim 8$ \AA) much larger than a single tetrahedron ($3.5$ \AA), at all temperatures measured, from 100 mK to 14 K.  Figure \ref{fig:rawdata} a) shows an elastic scan taken with $E_i = E_f$ = 13.5 meV at $T = 1.7$ K (after subtracting 14 K data), which reveals diffuse magnetic scattering throughout a larger range of $\mathbf{Q}$. Note in particular the absence of diffuse scattering at (222).  Figure \ref{fig:widths} b) also shows that the inelastic scattering takes on the same pattern in the thermal spin liquid phase (14 K) as it does in the frozen phase (100 mK) (``empty can'' background subtractions made in both panels).   In Fig. \ref{fig:rawdata} b and c) we show raw elastic scattering data (no subtraction) at $T = 1.7$ K and $T = 14$ K.

\section{Magnetization Data}
\label{sec:magnetization}
In order to investigate the possibility of an anisotropic $g$-tensor in \ncfo , magnetization measurements were performed at temperatures above the freezing transition ($T > 2.4$ K), using the extraction magnetometry technique in a commercial physical properties measurement system.   Measurements with the field applied along three different crystallographic axes were compared.  The data taken at $T= 40$ K are shown in Fig. \ref{fig:mag} a) and b).  Only slight deviations from isotropic behavior are observed at the highest field strengths ($\sim$ 9 T), and these could easily be due to demagnetization effects for crystals having slightly different shapes for the different field orientations.  Furthermore, the deviation from isotropic magnetization does not correspond to the expected hierarchy for either $XY$-like ($|M_{(110)}| > |M_{(111)}| > |M_{(100)}|$) or Ising-like ($|M_{(100)}| > |M_{(111)}| > |M_{(110)}|$) $g$-tensors.  For example, the magnetization of an ideal pyrochlore paramagnet with an $XY$-like $g$-tensor is shown in Fig. \ref{fig:mag} c), using the equation,
\begin{equation}
M_{\mathbf{d}}(H,T) = g_{\mathbf{d}} J \mu_B B_J(g_{\mathbf{d}} \mu_B J H / k_B T),
\label{eqn:brillouin}
\end{equation}

where $H$ is the applied magnetic field, $g_{\mathbf{d}}$ is the average projection of the $g$-tensor onto the field direction (averaged over the four sites on the tetrahedron), $J$ is the effective angular momentum, here taken to be $1/2$ since we may assume a spin-orbit coupled Kramers doublet ground state for Co$^{2+}$, and $B_J$ is the Brillouin function.  In Fig. \ref{fig:mag} c) an $XY$-like $g$-tensor was assumed, with $g_{xy} = 3.6$ and $g_{z} = 3.0$.  Although this equation is not expected to be valid at $T= 40$ K for \ncfo, since $T<\theta_{CW}$, one may expect the same hierarchy of magnetization strengths to be observed, even in such a correlated paramagnetic regime.   Thus, at least to within the demagnetization effects in these measurements, the $g$-tensor anisotropy in \ncfo \ is shown to be small on average.

%

\end{document}